\documentclass[10pt,conference]{IEEEtran}
\IEEEoverridecommandlockouts
\usepackage{cite}
\usepackage{amsmath,amssymb,amsfonts}
\usepackage{algorithmic}
\usepackage{graphicx}
\usepackage{textcomp}
\usepackage{xcolor,colortbl}
\usepackage{ifthen}
\usepackage{xspace}
\usepackage{booktabs}
\usepackage{listings}
\usepackage{algorithmic}
\usepackage{graphicx}
\usepackage{textcomp}
\usepackage[bookmarks=false,hidelinks]{hyperref}
\usepackage{soul}
\usepackage[most]{tcolorbox}
\usepackage{balance}
\usepackage{makecell}
\usepackage{caption}
\usepackage{listings}
\usepackage{multirow}
\usepackage{array}
\usepackage{setspace}

\def\BibTeX{{\rm B\kern-.05em{\sc i\kern-.025em b}\kern-.08em
    T\kern-.1667em\lower.7ex\hbox{E}\kern-.125emX}}
\begin{document}

\newcommand{\ie}{\emph{i.e.,}\xspace}
\newcommand{\eg}{\emph{e.g.,}\xspace}
\newcommand{\etc}{etc.\xspace}
\newcommand{\etal}{\emph{et~al.}\xspace}
\newcommand{\secref}[1]{Section~\ref{#1}\xspace}
\newcommand{\figref}[1]{Fig.~\ref{#1}\xspace}
\newcommand{\listref}[1]{Listing~\ref{#1}\xspace}
\newcommand{\tabref}[1]{Table~\ref{#1}\xspace}
\newcommand{\algoref}[1]{Algorithm~\ref{#1}\xspace}
\newcommand{\eqqref}[1]{Equation~(\ref{#1})\xspace}
\newcommand{\tool}[1]{{\sc #1}\xspace}
\newcommand{\vda}{$\hat{A}_{12}$\xspace}
\newcommand*\circled[1]{\tikz[baseline=(char.base)]{
  \node[shape=circle,draw,inner sep=1pt] (char) {#1};}}
\newcommand{\unknown}{\textcolor{red}{XXX}\xspace}
\newcommand{\mysubsubsection}[1]{\medskip\noindent\textbf{#1}.\xspace}
\newcommand{\subsubsubsection}[1]{\smallskip\noindent\underline{\textit{#1}}.\xspace}

\newcommand\FEDE[1]{\textcolor{cyan}{FEDERICO\texttt{->}#1}}
\newcommand\VIC[1]{\textcolor{purple}{VITTORIO\texttt{->}#1}}
\newcommand\LUCA[1]{\textcolor{teal}{LUCA\texttt{->}#1}}
\newcommand\TMPPAR[1]{\textcolor{blue}{#1}}

\title{Investigating Execution-Aware Language Models for Code Optimization\thanks{This article has been accepted for publication in \emph{The 33rd IEEE/ACM International Conference on Program Comprehension (ICPC 2025), RENE Track}.
This version of the manuscript is a preprint and may differ from the final published
version in terms of content and formatting.
}}
 \author{\IEEEauthorblockN{1\textsuperscript{st} Given Name Surname}
 \IEEEauthorblockA{\textit{dept. name of organization (of Aff.)} \\
 \textit{name of organization (of Aff.)}\\
 City, Country \\
 email address or ORCID}
 \and
 \IEEEauthorblockN{2\textsuperscript{nd} Given Name Surname}
 \IEEEauthorblockA{\textit{dept. name of organization (of Aff.)} \\
 \textit{name of organization (of Aff.)}\\
 City, Country \\
 email address or ORCID}
 \and
 \IEEEauthorblockN{3\textsuperscript{rd} Given Name Surname}
 \IEEEauthorblockA{\textit{dept. name of organization (of Aff.)} \\
 \textit{name of organization (of Aff.)}\\
 City, Country \\
 email address or ORCID}
 \and
 \IEEEauthorblockN{4\textsuperscript{th} Given Name Surname}
 \IEEEauthorblockA{\textit{dept. name of organization (of Aff.)} \\
 \textit{name of organization (of Aff.)}\\
 City, Country \\
 email address or ORCID}
 \and
 \IEEEauthorblockN{5\textsuperscript{th} Given Name Surname}
 \IEEEauthorblockA{\textit{dept. name of organization (of Aff.)} \\
 \textit{name of organization (of Aff.)}\\
 City, Country \\
 email address or ORCID}
 \and
 \IEEEauthorblockN{6\textsuperscript{th} Given Name Surname}
 \IEEEauthorblockA{\textit{dept. name of organization (of Aff.)} \\
 \textit{name of organization (of Aff.)}\\
 City, Country \\
 email address or ORCID}
 }
\author{\IEEEauthorblockN{Federico {Di Menna}\textsuperscript{\textdagger}, Luca Traini\textsuperscript{\textdagger}, Gabriele {Bavota}\textsuperscript{\textdaggerdbl}, Vittorio Cortellessa\textsuperscript{\textdagger}}
\IEEEauthorblockA{\textit{\textsuperscript{\textdagger}University of L’Aquila, L'Aquila, Italy}\\
\textit{\textsuperscript{\textdaggerdbl}Software Institute - Università della Svizzera Italiana, Switzerland}\\
 federico.dimenna@graduate.univaq.it, luca.traini@univaq.it, gabriele.bavota@usi.ch, vittorio.cortellessa@univaq.it}
}

\maketitle


\definecolor{pblue}{rgb}{0.13,0.13,1}
\definecolor{pgreen}{rgb}{0,0.4,0}
\definecolor{pred}{rgb}{0.9,0,0}
\definecolor{pgrey}{rgb}{0.46,0.45,0.48}
\definecolor{ppurple}{rgb}{0.58,0,0.82}

\lstdefinestyle{c_style}{   
  language=C++,
  showspaces=false,
  showtabs=false,
  keepspaces=true,
  breaklines=true,
  showstringspaces=false,
  breakatwhitespace=true,
  commentstyle=\color{pgreen}\ttfamily\itshape, 
  keywordstyle=\color{pblue}\bfseries,         
  stringstyle=\color{pred},                    
  upquote=true,
  basicstyle=\ttfamily\footnotesize,
  columns=fullflexible,
  frame=tb,
  aboveskip=1em,
  belowskip=1em,
  lineskip=3pt,
  morekeywords={},
  identifierstyle=\color{black},                     
  directivestyle=\color{ppurple}\bfseries,           
  emphstyle={\color{pblue}\bfseries},                
}

\begin{abstract}
Code optimization is the process of enhancing code efficiency, while preserving its intended functionality. This process often requires a deep understanding of the code execution behavior at run-time to identify and address inefficiencies effectively.
Recent studies have shown that language models can play a significant role in automating code optimization. However, these models may have insufficient knowledge of how code execute at run-time.
To address this limitation, researchers have developed strategies that integrate code execution information into language models. These strategies have shown promise, enhancing the effectiveness of language models in various software engineering tasks.
However, despite the close relationship between code execution behavior and efficiency, the specific impact of these strategies on code optimization remains largely unexplored.
This study investigates how incorporating code execution information into language models affects their ability to optimize code. Specifically, we apply three different training strategies to incorporate four code execution aspects --- line executions, line coverage, branch coverage, and variable states --- into CodeT5+, a well-known language model for code.
Our results indicate that execution-aware models provide limited benefits compared to the standard CodeT5+ model in optimizing code.
\end{abstract}


\label{sec:intro}
\section{Introduction}

Performance is a critical quality attribute of modern software systems \cite{Eaton2012}. Improving software performance involves activities across various layers of the software stack, ranging from architectural design \cite{Aleti2013} to compiler optimization \cite{Ashouri2018}. Among these, source code optimization --- the process of refining code by selecting efficient data structures and algorithms --- stands out as a key approach for enhancing the efficiency of software systems.

As of today, code optimization tasks largely rest on the shoulders of  developers. However, with the rising trend of using AI to automate software engineering tasks \cite{Hou2024, Fan2023, Sun2025, Ding2025, Xiao2024, Traini2024}, recent studies have begun exploring the potential of language models to automate code optimization \cite{shypula2024, gao2024, garg2024, Garg2022}. In these studies, language models are provided with a non-optimized version of a code snippet (e.g., a function) and tasked with generating an optimized version that improves specific performance properties, such as execution time.

While these efforts have demonstrated the significant potential of language models in automating code optimization \cite{shypula2024,gao2024, garg2024, Garg2022}, these models are typically trained on a ``static'' source code representation and may therefore lack an understanding of how code executes at run-time \cite{ma2024lmsunderstandingcodesyntax, Ni2024, Ding2024SemCoder}.
 Recent studies have shown that integrating these models with code execution information can significantly enhance their effectiveness across a variety of downstream software engineering tasks \cite{Ni2024, Ding2024SemCoder, huang-etal-2024-code, Ding2024Traced}.
For instance, Ding \etal \cite{Ding2024Traced} proposed a pre-training strategy to teach language models specific aspects of code execution, such as branch coverage and variable states, demonstrating improvements in tasks like clone retrieval and vulnerability detection. Similarly, Ni \etal \cite{Ni2024} introduced \emph{NExT}, a method that enables language models to inspect variable states of executed code lines and reason about their execution behavior, resulting in a higher fix rate for program repair tasks.
Despite these and other efforts \cite{Ding2024SemCoder, huang-etal-2024-code, Souza2023,mammadov2024learningprogrambehavioralmodels,chen2024}, the impact of execution-awareness in automated code optimization remains largely unexplored.

Given the close relationship between run-time execution behavior and code efficiency, this paper investigates how teaching language models to understand code execution behavior affects their effectiveness in optimizing code.
Specifically, we first train a CodeT5+ model \cite{wang2023codet5p} with training objectives related to four code execution aspects, namely number of line executions, line coverage, branch coverage, and variable states. We then evaluate and compare the code speed-ups achieved by these execution-aware models against those of the standard CodeT5+ model.
Our findings reveal that execution-aware models do not outperform the traditional language model in code optimization. On the contrary, they seem to be less effective than the standard CodeT5+ model.

The contribution of this paper are as follows: 
\begin{itemize}
	\item A comprehensive evaluation of twelve execution-aware language models, based on CodeT5+, for code optimization. This evaluation encompasses four different code execution aspects and three distinct strategies.
	\item A complementary analysis providing insights to guide future research in automated code optimization.
	\item A replication package to reproduce our findings\footnote{\url{https://github.com/SpencerLabAQ/exec-aware-code-opt}}.
\end{itemize}

\paragraph*{Paper Structure} \secref{sec:background} provides background on automated code optimization and execution-aware language models for code. \secref{sec:design} describes our study design and research questions. \secref{sec:setup} explains the experimental setup. \secref{sec:results} reports the results. \secref{sec:discussion} presents a qualitative analysis, discusses the results, and describes threats to validity. \secref{sec:conclusion} concludes this paper.

\section{Background and Related Work}
\label{sec:background}

\subsection{Language Models for Code Optimization}

Transformer-based language models are a category of deep learning models that have driven significant advancements in natural language processing. In recent years, studies have demonstrated the ability of such models to effectively perform various software engineering tasks \cite{Hou2024, Fan2023, Sun2025, Ding2025,Xiao2024, Mastropaolo2024, Mastropaolo2024ICSE, Sejfia2024, Huang2024, Akhilesh2023}. 
Since code optimization involves generating a refinement of a source code that improves efficiency, it can be directly framed as a code-to-code transformation task (specifically, from inefficient to efficient version). 
For example, Garg \etal proposed \textit{DeepDev-PERF} \cite{Garg2022}, a transformer-based model pretrained on English and C\# source code corpora, and further specialized with performance-improving commits. Their work shows that \textit{DeepDev-PERF} can generate
code changes that lead to tangible performance optimizations to various open source C\# projects.

Shypula \etal proposed a dataset of performance-improving code edits (\textit{PIE}) \cite{shypula2024}, composed of slow-fast C++ program pairs. They exploit this dataset to evaluate a variety of techniques, including fine-tuning and prompting, for adapting well-known language models, such CodeLLama, GPT-3.5 and GPT-4, for code optimization.
Their results show that certain combinations of techniques and models can achieve higher optimizations than human reference.
Similarly, Chen \etal \cite{Chen2024Supersonic} proposes \textit{SUPERSONIC}, a smaller seq2seq model targeting minor source code modifications for optimization. \textit{SUPERSONIC} is trained on C/C++ slow-fast program pairs leveraging a diff-based input representation.  Results show that it outperforms larger language models, such as GPT-3.5-Turbo and GPT-4, on code optimization while retaining significant similarity with the original program.  
RAPGen \cite{garg2024} leverages a pre-constructed knowledge-base to prompt language models in zero-shot to generate code inefficiencies fix.
Gao \etal \cite{gao2024} modeled the optimization task with a search-based approach, integrating language models with evolutionary search and outperforming several baselines models such as CodeLlama, Gemini and ChatGPT.

Although these studies have demonstrated the effectiveness of language models in optimizing code, they tend overlook the potential implications of integrating code execution information into the model.

\subsection{Execution-aware Language Models}
Most existing language models for code focus on learning static form of code text,  however they may lack a semantic understanding of how code execute at run-time \cite{ma2024lmsunderstandingcodesyntax, Ni2024}.
To mitigate this issue, several approaches have been proposed to integrate language models with  code execution information.
For instance, Ding \etal proposed \textit{TRACED} \cite{Ding2024Traced}, an execution-aware pre-training strategy to capture run-time execution aspects of code.
This strategy pre-trains the language model with the task objective of predicting variable states, and line coverage, forcing the model to reason about code execution behavior.
They found that learning code execution behavior, significantly improves the capabilities of language models in both clone retrieval and vulnerability detection tasks.

Another example is \textit{SemCoder} \cite{Ding2024SemCoder}, an execution-aware language model that leverages monologue reasoning to learn about local execution effects of individual statements, overall input/output behavior, thereby linking static code with code execution behavior. 
\textit{SemCoder} shows competitive effectiveness with GPT-3.5-turbo on code generation and execution reasoning tasks, despite being significantly (6.7B parameters).
A similar approach, namely \textit{NExT}, was recently proposed by Ni \etal \cite{Ni2024}. This method teaches the language models to inspect the execution traces of programs (variable states of executed lines) and reason about their run-time behavior through chain-of-thought rationales. 
\textit{NExT} considerably improves the fix rate of a PaLM 2 model for program repair tasks.

Huang \etal \cite{huang-etal-2024-code} proposed \textit{FuzzPretrain}, a method to explore
 run-time execution information of code revealed by their test cases and embed it into the feature
representations of code as complements.
\textit{FuzzPretrain} yields significant improvements on the code search task
over its language model counterparts trained with only static source code.

Previous research has studied the effectiveness of execution-aware models across various software engineering tasks, such as program repair \cite{Ni2024}, vulnerability detection \cite{Ding2024Traced}, clone retrieval \cite{Ding2024Traced}, code generation \cite{Ding2024SemCoder}, and code search \cite{huang-etal-2024-code}. However, despite the clear connection between runtime execution behavior and code efficiency, no prior work has directly examined how incorporating code execution information into language models influences their ability to optimize code. This paper seeks to address this gap.

\section{Study Design}\label{sec:design}

The \textit{goal} of this study is to investigate how integrating code execution information into language models impacts their ability to optimize code. We analyze twelve execution-aware models, derived from CodeT5+, that incorporate different combinations of four code execution aspects and three training strategies.
Specifically, for each code execution aspect, we create execution-aware models using three distinct training strategies. We then evaluate the code optimizations provided by these execution-aware models and compare them to those produced by the standard CodeT5+ model.

In the following sub-sections, we describe the code execution aspects and training strategies investigated in this study, as well as the research questions we aim to address.

\begin{figure}[t]
    \captionsetup{justification=centering}
    \center\includegraphics[width=.9\columnwidth]{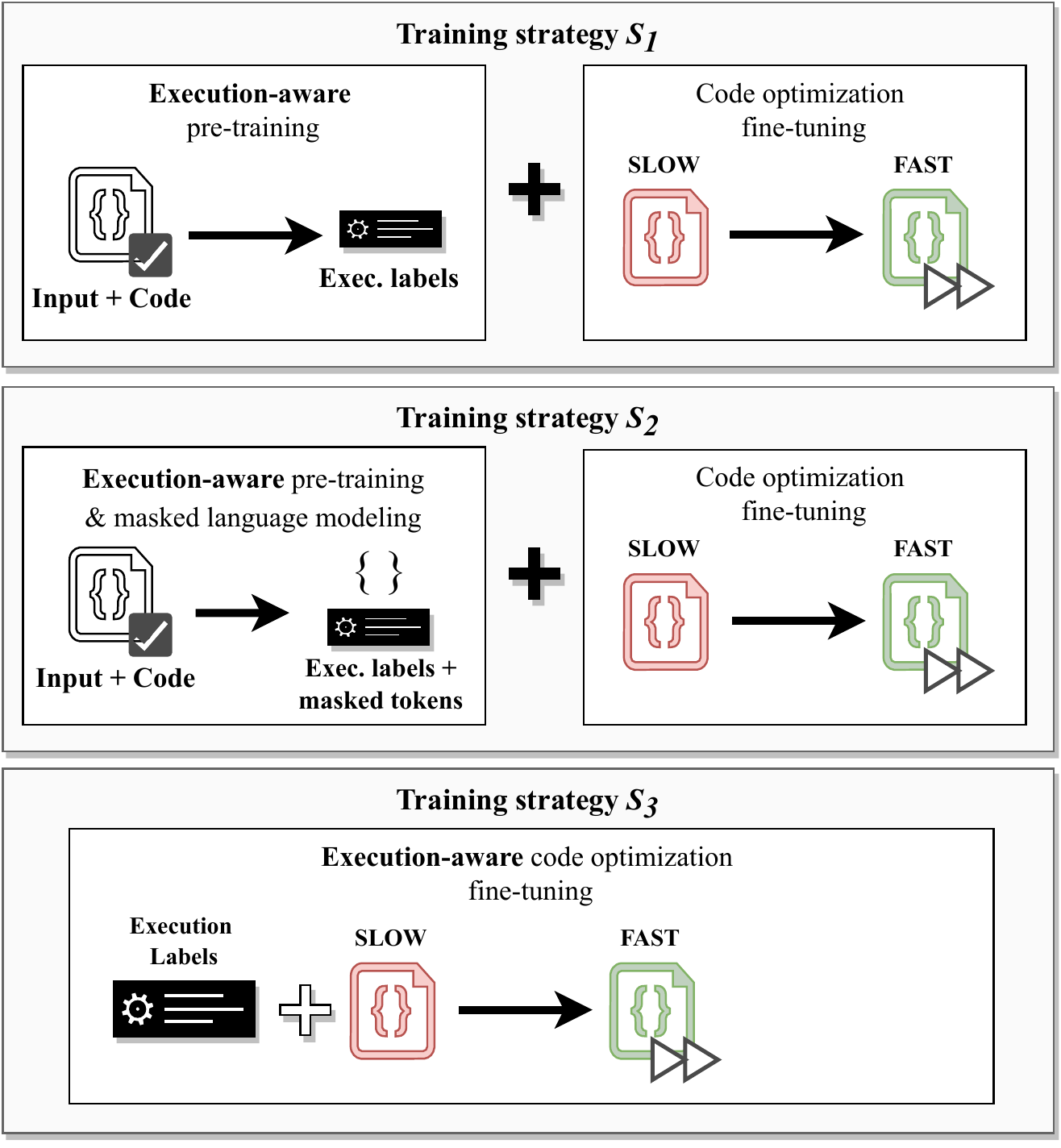}
    \caption{Overview of the proposed training strategies for building execution-aware language models. }\label{fig:approach}
\end{figure}

\subsection{Code Execution Aspects}

\subsubsection{Line Executions (LE)}
This aspect refers to how often a specific line of source code is executed during runtime.
Each line of code is assigned a label indicating the number of times it has been executed. This information provides understanding on \emph{hotspots} within the code, \ie code sections that are frequently executed and may benefit from optimization.

\subsubsection{Line Coverage (LC)}
Line coverage identifies which lines of source code are executed at least once during execution.
This provides insights about which parts of source code are covered after executing the program with a specific input, and which one remain underutilized.

\subsubsection{Branch Coverage (BC)}
This aspect is primarily designed to determine the parts of the source code associated to a branch. Given the execution of the program with a specific input, each line of code can fall into one of three categories: (i) it is part of a covered branch, (ii) it is part of an uncovered branch, or (iii) it is not part of any branch. 

\subsubsection{Variable States (VS)}
This information focuses on how the code execution changes the state of variables. For example, some variables may control the program’s flow (\eg counters or boolean flags), while others store intermediate or final results to be returned. Understanding the state of these variables at the end of execution provides insights into the code behavior at runtime.

\subsection{Training Strategies}
\label{subsec:strategies}
We define three distinct training strategies, each applied to a language model which has been already pre-trained on a static representation of code text, specifically CodeT5+.
Each strategy is designed to teach the model a particular aspect of code execution using a specific training methodology.
Figure \ref{fig:approach} presents a high-level overview of the stages involved in each training strategy.
In particular, we consider the following strategies: (i) \textit{execution-aware pre-training} ($S_1$), (ii) \textit{execution-aware pre-training} combined with \textit{masked language modeling} ($S_2$), and (iii) \textit{execution-aware fine-tuning} ($S_3$).
 In the following, we provide a description of each training strategy, while their instantiations to specific training datasets are detailed in Section \ref{sub:datasets}.

\smallskip
\textit{$S_1$ --- Execution-aware Pre-training + Fine-tuning}: 
This strategy involves two sequential training stages. In the first stage, the model undergoes pre-training to predict aspects of code execution based on specific input test cases. To deepen the model understanding of runtime code execution, we train it on multiple input cases to capture diverse execution behaviors of the same code. We term this process \textit{execution-aware pre-training}. Specifically, the model learns to predict run-time execution information without actually running the code (\ie it is provided with a program and a specific input and it must predict a specific execution aspect, like for example the line coverage). This step is designed to teach the language model how code executes at runtime.
In the second stage, the execution-aware model is fine-tuned for code optimization. This is framed as a sequence-to-sequence downstream task, where the input is a slower version of a program, and the output is an optimized, faster version. Importantly, the optimization process aims to preserve the program’s intended functionality; that is, the optimized program must produce the expected output.

\smallskip
\textit{$S_2$ --- Execution-aware Pre-training \& Masked Language Modeling + Fine-tuning}: Ding \etal~\cite{Ding2024Traced} demonstrated that combining execution-aware pre-training with \emph{masked language modeling} (MLM) can enhance model effectiveness across multiple software engineering tasks. Building on this approach, this strategy combines an execution-aware objective (as the one described for $S_1$) with an MLM objective during pre-training.
The MLM objective involves randomly masking 15\% of tokens in the input sequence and training the model to predict the masked tokens based on the surrounding context. 
Following the pre-training stage, as in the $S_1$ strategy, the model is fine-tuned for code optimization.

\smallskip
\textit{$S_3$ --- Execution-aware Fine-tuning}: This strategy does not involve a pre-training stage. Instead, code execution information is incorporated directly into the model input (\ie the slow version of the code) during fine-tuning. This information is added either line-by-line (for $LE$, $LC$, and $BC$) or at the end of the code (for $VS$).
Unlike the first two strategies, this one assumes the availability of an execution trace for the input code. Since code execution information is typically tied to a specific input test case, we adopt different methods depending on the aspect of code execution being considered.
For \emph{Line Executions} ($LE$), we use the maximum number of executions recorded for each line across all execution traces. This choice reflects the assumption that these maximum values are more impactful on execution time. For the other code execution aspects, we select a single trace corresponding to a random test case to annotate the slow code.

\subsection{Evaluation Metrics}
\label{subsec:metrics}
To evaluate the model effectiveness in code optimization, we use three evaluation metrics introduced in prior work \cite{shypula2024}:

\begin{itemize}
\item \emph{Correct (\%)}: The percentage of generated programs that successfully execute and produce the expected output for all available test cases.
\item \emph{Speedup}: The absolute improvement in terms of execution time. It is defined as  $Speedup=\frac{T_{I}}{T_{G}}$ , where  $T_{I}$  and  $T_{G}$ denote the execution times of the input and generated programs, respectively. In line with previous work \cite{shypula2024}, we set \emph{Speedup = $1$} when the generated programs are either incorrect or slower than the input program.
\item \emph{Percent Optimized (\%Opt)}: The percentage of generated code that is both correct and faster across the entire testing set. A generated program is considered faster if it achieves a speed improvement of at least 10\%, corresponding to a \emph{Speedup} of at least 1.1x.
\end{itemize}

\subsection{Research Questions}

This study aims to answer the following research questions:

\smallskip
\begin{enumerate}
    \item [\textbf{RQ$_{1}$}]\textit{How does learning line executions impact the effectiveness of language models for code optimization?} Several code optimizations aim to reduce the frequency of line executions (\eg loop optimization \cite{Song2017}). This research question investigates the impact of teaching a model to understand line execution behavior on its ability to optimize code.\smallskip
    \item [\textbf{RQ$_{2}$}]\textit{How does learning line coverage impact the effectiveness of language models for code optimization?} Here, we study how incorporating line coverage information into the language model affects its effectiveness in code optimization. \smallskip
    \item [\textbf{RQ$_{3}$}]\textit{How does learning branch coverage impact the effectiveness of language models for code optimization?} 
    Branch coverage information gives indications of how the conditional control flow statements impacts the code execution at run-time. This information can be crucial to identify potential optimization opportunities \cite{Nistor2015}. In this research question, we evaluate how learning such information influences the model effectiveness in optimizing code. \smallskip
    \item [\textbf{RQ$_{4}$}]\textit{How does learning variable states impact the effectiveness of language models for code optimization?} Having an understanding of the types and values of variables may provide meaningful insights into code execution behaviors, \eg variables elimination, type optimization and how the instructions reflects on the final values of used variables. Here we analyze whether teaching variable states to language models could enhance their ability to optimize code. 
    \smallskip
\end{enumerate}

To answer these RQs, we conducted an empirical study to evaluate the effectiveness of twelve execution-aware language models for code optimization. For each code execution aspect (\ie RQ), we developed three execution-aware models using the training strategies outlined in Section \ref{subsec:strategies}. We then evaluated the effectiveness of these models using the metrics defined in Section \ref{subsec:metrics} and compared to the standard CodeT5+ model.

\section{Experimental Setup}
\label{sec:setup}

In this section, we outline a detailed description of our experimental procedure, including the datasets construction, training procedure, baselines and models evaluation.

\subsection{Datasets Construction}
\label{sub:datasets}
Strategies $S_1$ and $S_2$ require an execution-aware pre-training dataset as well as a fine-tuning dataset, whereas the strategy $S_3$ only needs an execution-aware fine-tuning dataset. We start by describing the data related to the code optimization task and execution information, followed by an explanation of how this data is transformed into datasets tailored to each specific training strategy.

\smallskip
\subsubsection{Code optimization}
\label{sub:codeopt}
We train language models for code optimization using the \emph{Performance Improving Edits (PIE)} dataset \cite{shypula2024}, which consists of C++ slow-fast program pairs derived from from IBM \emph{CodeNet} \cite{puri2021codenet}.
\emph{CodeNet} is a large-scale dataset of programs, with each program representing a solution to a specific coding problem. It includes multiple programs for each coding problem, which may be authored either by the same developer or by different developers. Additionally, it provides test cases that can be used to evaluate the correctness of each program (based on the problem specifications) and measure its execution time. 
The \emph{PIE} dataset involve 77k `\emph{slow}-\emph{fast}' program pairs, where both programs in each pair aim to solve the same coding problem. \emph{PIE} is already provided in a pre-split format consisting of train/validation/test subsets, ensuring that programs belonging to each problem only appear in one of those subsets. Additionally, we apply some preprocessing steps to canonicalize the code and remove any subjective elements in comments and coding style, as outlined in previous work \cite{Chen2024Supersonic}. We use the \emph{gcc} preprocessor to remove comments and \emph{clang-format} to format programs based on the LLVM coding style. 

\begin{figure*}[t]
    \captionsetup{justification=centering}
    \center\includegraphics[width=\textwidth]{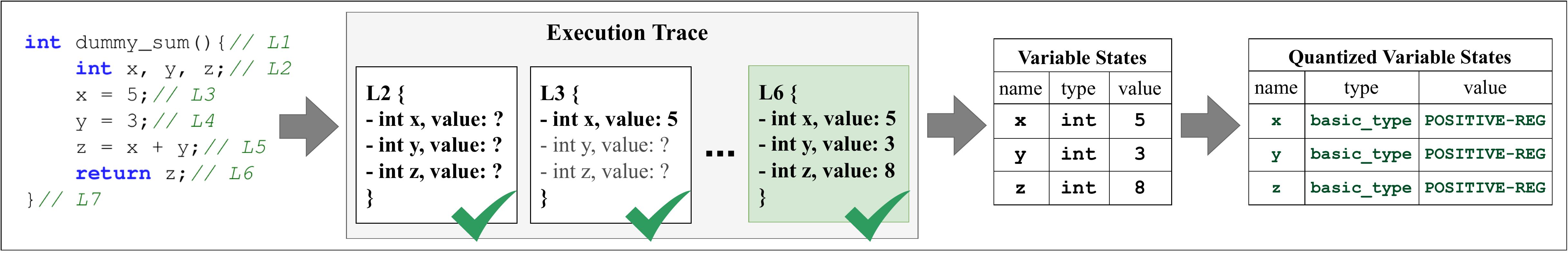}
    \caption{The figure presents the execution trace corresponding to the outlined sample function, namely \texttt{dummy\_sum()}. We use the question mark symbol (?) to show that a variable does not yet have an assigned value. Additionally, it also depict the \emph{variable states} information and the associated quantized values. }\label{fig:trace_vars}
\end{figure*}

\smallskip
\subsubsection{Execution traces}
To obtain code execution information, we use the \textit{gdb}\footnote{\url{https://sourceware.org/gdb/}} debugger to trace program execution, following a method similar to that used in \cite{Ding2024Traced}. To prevent prolonged computations, the tracing duration was limited to 500 seconds. The execution tracing procedure may fail if it takes too long or causes an invalid trace due to an unexpected execution interruption (\eg program crash). 
The output of the tracing procedure consists of program execution traces that document the complete execution history of each program run. Specifically, such execution traces list all the executed lines, alongside with the name, type, and value of the corresponding program variables at each step of the program execution.
We then preprocess these traces to extract execution information needed for our study, such as the execution frequency of each line of code and the state of variables at the end of program execution. For sake of clarity, we depict a sample execution trace referring to a dummy program in Figure \ref{fig:trace_vars}. As illustrated, the trace displays the name, type and value of each variable for every line of code that is run, represented as blocks. The last block in the trace (highlighted in green in the figure) contains the final variable information we use for the \emph{variables state} execution aspect. 

\subsubsection{Datasets}
For fine-tuning the model for code optimization, we directly rely on the \emph{PIE} dataset for both the $S_1$ and $S_2$ strategies. Conversely, in execution-aware fine-tuning (\ie $S_3$), the dataset must be adapted to incorporate execution information into the \emph{slow} input program.
To build the $S_3$ execution-aware fine-tuning dataset, we performed the tracing procedure over all the programs included in the `\emph{slow}' column of \emph{PIE} dataset. Note that, a `\emph{slow}' program can appear multiple times in the \emph{PIE} training dataset if associated to multiple faster versions. 
Specifically, among the 21,713 distinct `\emph{slow}' programs contained in the \emph{PIE} training set, 9,145 programs were successfully traced. Similarly, from 978 `\emph{slow}' programs contained in the test set, 466 were traced successfully. 
This resulted in a total of 31,585 slow-fast training pairs and 466 instances for the testing set. 

Differently, to construct the execution-aware pre-training dataset (for $S_1$ and $S_2$), we initially excluded all the C++ \emph{CodeNet} programs appearing in either the `\emph{slow}' or `\emph{fast}' columns of the entire \emph{PIE} dataset (43,606 programs). This filter prevents duplicate samples present in both the pre-training and fine-tuning datasets. 
Additionally, for each coding problem in \emph{CodeNet}, we randomly select up to 150 program-test cases pairs, and we collect their corresponding execution traces.
This approach ensures traces collection in a reasonable timeframe, as \emph{CodeNet} comprises more than 8 millions of C++ programs. Indeed, the tracing collection procedure resulted in 119,764 execution traces for building the execution-aware pre-training dataset, involving 87,705 distinct programs. It is important to note that the same program may appear multiple times in this dataset if the corresponding coding problem involves multiples test cases, as each test case generates distinct execution traces.

\subsection{Quantization Strategy}
Language models may face challenges in learning code execution information, as they are primarily continuous numerical values \cite{Ding2024Traced}.
Let us consider the example of line execution, where each line is associated with the number of times it has been executed for a specific input. This value can be any integer greater than or equal to zero.
Thus, it may be challenging for the model to learn the underlying patterns behind this code execution aspect.
To address this, as proposed in prior work \cite{Ding2024Traced}, we define a set of special tokens for each execution aspect to quantize the concrete execution information.
For \emph{line executions}, we performed a quantization of concrete values based on specific ranges, resulting in four unique tokens; \texttt{<e>}: $1$ execution, \texttt{<e+>}: $2-5$ executions, \texttt{<E>}: $6-20$ executions, \texttt{<E+>}: $21+$ executions. Lines that are not executed are not marked with any token. The majority of program lines were observed to have zero or one single execution, representing over 80\% of all analyzed lines. In light of this, we allocated a dedicated token (\texttt{<e>}) for representing a single execution case. We determined the other thresholds by studying how often lines of code were executed more than once. In particular, we set the higher threshold using the outlier detection rule \hbox{\emph{T = Q3 + 2.5 * IQR}}, and subsequently divided the remaining values into two groups, using the median value as the dividing criterion. The calculated values for median and \emph{T} were 4 and 23, which we rounded to 5 and 20. 
To identify the executed lines in the \emph{line coverage} aspect, we only re-use the \texttt{<e>} token; uncovered lines remain unlabeled, as for the $LE$ aspect. 
For learning \emph{branch coverage} we use two tokens, namely \texttt{<BC>} and \texttt{<BNC>}, to indicate that a line is part of a branch that is covered or not covered, respectively. In this case, lines that do not belong to any branch are left unlabeled. 
Regarding \emph{variable states}, we adapt the quantized representation introduced in TRACED \cite{Ding2024Traced} to represent C++ variables, using their name, type and value. 
In Figure \ref{fig:trace_vars}, we report the variable states quantization step for a dummy example. As depicted, the \texttt{int} became \texttt{basic\_type}, while the concrete values have been converted into \texttt{POSITIVE-REG}. We outline all details related to the quantization strategy for each execution aspect in our replication package description.

In Table~\ref{tab:code_ex_example}, we provide a comprehensive example of the execution information that we feed into the language models. In the first column, we report the traced program alongside with a test case (\ie program input) and the output obtained from its execution. The other columns outline the special tokens computed for each execution aspect. As show in the table, the \emph{variable states} information refers the entire code execution, whereas the other aspects correspond to line-by-line execution information.

\lstset{basicstyle=\small,style=c_style}


\renewcommand{\arraystretch}{1.1} 

\begin{table*}[!ht]
    \centering
    \caption{Sample C++ source code aligned with the corrisponding execution information.}
    \scriptsize
    \begin{tabular}{l>{\centering\arraybackslash}c>{\centering\arraybackslash}c>{\centering\arraybackslash}c>{\centering\arraybackslash}c>{\centering\arraybackslash}l>{\centering\arraybackslash}l}
        \toprule
        \multirow{2}*{\textbf{C++ Source Code}} & \multirow{2}*{\textbf{\makecell{Line \\Exec.}}} & \multirow{2}*{\textbf{\makecell{Line \\Coverage}}} & \multirow{2}*{\textbf{\makecell{Branch \\Coverage}}} & \multicolumn{3}{c}{\textbf{Variable States}} \\ \cline{5-7}
            & & & & \multicolumn{1}{c}{\textbf{Var. Name}} & \multicolumn{1}{c}{\textbf{Var. Type}} & \multicolumn{1}{c}{\textbf{Quantized Val.}} \\ \hline
           \textbf{Test Case}: \texttt{keyofscience} & & & & & &\\ \hline 
            \lstinline!#include<iostream>! & - & - & - & \texttt{k} & \texttt{class} & \texttt{OTHER} \\ 
           \lstinline!#include<string>! & - & - & - & \texttt{S} & \texttt{class} & \texttt{OTHER} \\ 
           \lstinline!using namespace std;! & - & - & - & \texttt{s} & \texttt{basic\_type} & \texttt{POSITIVE-REG} \\ 
           \lstinline!int main(){! & \texttt{<e>} & \texttt{<e>} & - & \texttt{f} & \texttt{basic\_type} & \texttt{POSITIVE-REG} \\ 
           \lstinline!    int s=0,f=0;! & \texttt{<e>} & \texttt{<e>} & - & \texttt{i} & \texttt{basic\_type} & \texttt{POSITIVE-REG} \\ 
           \lstinline!    string S,k="keyence";! & \texttt{<e+>} & \texttt{<e>} & - & \cellcolor{blue!3}~ & \cellcolor{blue!3}~ & \cellcolor{blue!3}~ \\ 
           \lstinline!    cin>>S;! & \texttt{<e>} & \texttt{<e>} & - & \cellcolor{blue!3}~ & \cellcolor{blue!3}~ & \cellcolor{blue!3}~ \\ 
           \lstinline!    for(int i=0;i<S.length();i++){! & \texttt{<e+>} & \texttt{<e>} & \texttt{<BC>} & \cellcolor{blue!3}~ & \cellcolor{blue!3}~ & \cellcolor{blue!3}~ \\ 
           \lstinline!        if(S[i]==k[i])s++;! & \texttt{<e+>} & \texttt{<e>} & \texttt{<BC>} & \cellcolor{blue!3}~ & \cellcolor{blue!3}~ & \cellcolor{blue!3}~ \\ 
           \lstinline!        else break;! & \texttt{<e>} & \texttt{<e>} & \texttt{<BC>} & \cellcolor{blue!3}~ & \cellcolor{blue!3}~ & \cellcolor{blue!3}~ \\ 
            \lstinline!    }! & - & - & - & \cellcolor{blue!3}~ & \cellcolor{blue!3}~ & \cellcolor{blue!3}~ \\ 
            \lstinline!    for(int i=0;i<S.length();i++){! & \texttt{<e+>} & \texttt{<e>} & \texttt{<BC>} & \cellcolor{blue!3}~ & \cellcolor{blue!3}~ & \cellcolor{blue!3}~ \\ 
            \lstinline!        if(S[S.length()-1-i]==k[6-i])f++;! & \texttt{<e+>} & \texttt{<e>} & \texttt{<BC>} & \cellcolor{blue!3}~ & \cellcolor{blue!3}~ & \cellcolor{blue!3}~ \\ 
            \lstinline!        else break;! & \texttt{<e>} & \texttt{<e>} & \texttt{<BC>} & \cellcolor{blue!3}~ & \cellcolor{blue!3}~ & \cellcolor{blue!3}~ \\ 
            \lstinline!    }! & - & - & - & \cellcolor{blue!3}~ & \cellcolor{blue!3}~ & \cellcolor{blue!3}~ \\ 
            \lstinline!    if(s+f>=7)cout<<"YES"<<endl;! & \texttt{<e>} & \texttt{<e>} & \texttt{<BC>} & \cellcolor{blue!3}~ & \cellcolor{blue!3}~ & \cellcolor{blue!3}~ \\ 
            \lstinline!    else cout<<"NO"<<endl;! & - & - & \texttt{<BNC>} & \cellcolor{blue!3}~ & \cellcolor{blue!3}~ & \cellcolor{blue!3}~ \\ 
            \lstinline!    return 0;! & \texttt{<e>} & \texttt{<e>} & - & \cellcolor{blue!3}~ & \cellcolor{blue!3}~ & \cellcolor{blue!3}~ \\ 
            \lstinline!}! & \texttt{<e>} & \texttt{<e>} & - & \cellcolor{blue!3}~ & \cellcolor{blue!3}~ & \cellcolor{blue!3}~ \\ \midrule
            \textbf{Output}: \texttt{YES} & & & & & &\\ \bottomrule
             \\ 

    \end{tabular}
    
    \label{tab:code_ex_example}
\end{table*}

\subsection{Training Procedure}

We exploit a Text-To-Text Transfer Transformer (T5) model, namely CodeT5+~\cite{wang2023codet5p}, which is specifically pre-trained on source code and widely adopted in software engineering literature \cite{Ahmed24,Luo24,Yu24,MastropaoloEASE24}. With regard to the available computational resources, we selected the CodeT5+ model with a parameter size of 220M. 
We ran all the experiments using the HuggingFace Transformers library. 
For all the tasks, we train the model for $1$ epoch, with a batch size of $16$ and a learning rate of \(1\mathrm{e}{-5}\). 
All experiments were conducted using the default AdamW optimizer. The fine-tuning script included in the official CodeT5+ release specifies the maximum token limits for the source and target inputs as $320$ and $128$, respectively\footnote{\url{https://github.com/salesforce/CodeT5/tree/main/CodeT5\%2B}}. Since the majority of our instances exceeds these threshold, we decided to increase both the source and target limits to $512$ tokens.
In Table \ref{tab:ds_strat}, we report the overall instances for each dataset resulting from filtering out the training samples with a number of tokens greater than $512$.

\begin{table}[t]
    \captionsetup{justification=centering}
    \center
    \caption{Overview of the total number of instances in the datasets after applying the maximum token limits filter. Note that different execution aspects can lead to dataset sizes, as the number of tokens may be different. }
    \resizebox{\columnwidth}{!}{ 
        \begin{tabular}{l|c|ccccc}
            \toprule
            \textbf{Dataset} & \textbf{BL} & \textbf{LE} & \textbf{LC} & \textbf{BC} & \textbf{VS} \\ 
            \midrule
            Execution-aware Pre-training &  - & 96,308 & 96,463 & 96,463 & 96,468 \\ \midrule
            Fine-tuning &  52,471 & \multicolumn{4}{c}{52,471} \\
            Execution-aware Fine-tuning &  28,753 & 28,654 & 28,244 & 28,654 & 28,034 \\
            \bottomrule
        \end{tabular}
    }
    
    \label{tab:ds_strat}
\end{table}

In the following, we outline our implementation of each training strategy.

\smallskip
\subsubsection{Strategy $S_1$}
Concerning the $S_1$ execution-aware pre-training, the language model reads a program code and an associated test case (\ie program input) and generates the corresponding execution information. Let us define the set of the program lines as \hbox{\textit{$L$ = \{l$_1$,...,l$_n$\}}} and the test case as $input$; the model prompt is then built as \hbox{\textit{$I_{PT\_S_1}$ = \{\texttt{classify:} + $input$ + \texttt{<SEP>} + l$_1$,...,l$_n$\}}}.
The corresponding target sequence contains the execution information tokens joined together, separated by the ``\texttt{\textbackslash n}" character.
Concerning line-wise execution information (\ie $LE$, $LC$ and $BC$), we can define the set of execution information tokens corresponding to the program \emph{L} as \hbox{\textit{$E$ = \{e$_1$,...,e$_n$\}}} and accordingly build the target sequence as \hbox{\textit{$O_{PT\_S1}$ = \{e$_1$,...,e$_n$\}}}.
For instance, assuming to deal with \emph{line executions} data, the output sequence corresponding to the example reported in Table \ref{tab:code_ex_example} would be encoded as \hbox{"\texttt{\textbackslash n\textbackslash n\textbackslash n<e>\textbackslash n<e>\textbackslash n<e+>\textbackslash n<e>...<e>\textbackslash n<e>}"}. For the \emph{variable states}, the target sequence consists of a list of inline code comments detailing the name, type, and value of each program variable. For instance, the code reported in Table \ref{tab:code_ex_example} would generate five comments to append to the code, in the following way: "\texttt{...// k class OTHER\textbackslash n...\textbackslash ni basic\_type POSITIVE-REG}".
\textit{$I_{PT\_S_1}$} and \textit{$O_{PT\_S_1}$} are fed into the language model as source and target sequences in a sequence-to-sequence supervised learning setting. 

For the $S_1$ fine-tuning, the code optimization task is modeled as a sequence-to-sequence downstream tasks, in which the model reads the `\emph{slow}' code and generates the optimized version. Assuming to have a slow program composed of \emph{n} lines (\hbox{\textit{$S$ = \{s$_1$,...,s$_n$\}}}) and a faster version composed of \emph{m} lines (\hbox{\textit{$F$ = \{f$_1$,...,f$_m$\}}}), the input sequence for the fine-tuning task is formulated as \hbox{\textit{$I_{FT\_S_1}$ = \{\texttt{optimize: } + s$_1$,...,s$_n$\}}}. The target sequence simply consists of the set of lines of the optimized version joined together: \hbox{\textit{$O_{FT\_S_1}$ = \{f$_1$,...,f$_m$\}}}. The `\texttt{classify:}' and `\texttt{optimize:}' prefixes are utilized to assist the language model in distinguishing between the specified tasks.

\smallskip
\subsubsection{Strategy $S_2$}
This strategy is similar to strategy $S_1$, but it integrates the \emph{masked language modeling} (MLM) objective to the execution-aware pre-training process.
The model alternates the batches of the execution-aware pre-training dataset for performing both execution-aware and MLM pre-training objectives, aligned with prior research~\cite{huang2023}. To distinguish between the two tasks, we use the `\texttt{mlm}:' input prefix for MLM batches instances.
Formally, let \hbox{\textit{$M$ = \{m$_1$,...,m$_k$\}}} representing the set of \emph{k} tokens encoding the input program, including those that are masked. 
The input sequence for MLM is then defined as \hbox{\textit{$I_{PT\_S_2}$ = \{\texttt{mlm:} + m$_1$,...,m$_k$\}}}, with the output being the sequence of real tokens corresponding to the masked ones. The instances contained in the batches that are not used for MLM are handled identically to $S_1$.
With this feeding process, we aim to teach the language models to generate both execution and MLM labels. The fine-tuning stage is performed identically to $S_1$, meaning that \hbox{\textit{$I_{FT\_S_2}$ = $I_{FT\_S_1}$}} and \hbox{\textit{$O_{FT\_S_2}$ = $O_{FT\_S_1}$}}.

\subsubsection{Strategy $S_3$}
The strategy $S_3$ relies to a direct execution-aware fine-tuning process, in which the tokens related to execution information are directly incorporated within the \emph{slow} code, by using code comments. 
For instance, the 8th line of code in Table~\ref{tab:code_ex_example} would be converted into the line \hbox{\lstinline!for(int i=0;i<S.length();i++){ // <e+>!} in case of \emph{line executions}. 
For \emph{line executions}, \emph{line coverage} and \emph{branch coverage} aspects, the execution information is generated for each line of code. Then, formalizing the set of source code lines annotated with executions tokens as \textit{$S_{E}$ = \{se$_1$,...,se$_n$\}}, the input sequence ($I_{FT\_S_3}$) is constituted by the flattened version of $S_{E}$. 
Differently, for the \emph{variable states} aspect we generate inline comments for each variable, similarly to $S_1$, and append them to the '\emph{slow}' program. The target sequence \textit{$O_{FT\_S_3}$} corresponds to the optimized version of the code, as for the other strategies. 

\smallskip
\subsection{Baselines}
We compare each execution-aware language model against the plain CodeT5+ model directly fine-tuned for the code optimization task (\ie without injecting any execution information). 

In that, we define the following baseline models: 

\begin{itemize}
    \item \textit{Baseline for $S_1$-$S_2$ ($BL_{S_{12}}$)}: We train this baseline using the fine-tuning dataset outlined in Table \ref{tab:ds_strat}, involving 52,471 overall instances. Note that the dataset used for training the baseline is identical to the dataset used for fine-tuning the execution-aware models with $S_1$ and $S_2$.
    \item \textit{Baseline for $S_3$ ($BL_{S_{3}}$)}: To guarantee a fair comparison regarding strategy $S_3$, we restrict the analysis to the subset of programs that were successfully traced, applying the same subset for training the baseline, leading to a total of 28,753 instances.  
\end{itemize}

\subsection{Models Evaluation}
We evaluate the effectiveness of each language model using the \emph{PIE} test set \cite{shypula2024}. Specifically, we prompt each model to generate an optimized version of each `\emph{slow}' program contained in the test set.
Subsequently, the `\emph{slow}' programs and the generated programs are compiled and simulated using the available \emph{CodeNet} test cases on the \emph{gem5} simulator \cite{Binkert2011}. Through this process, we assess the correctness of the generated programs and measure their execution times, following the methodology outlined in previous work \cite{shypula2024}.
The \emph{Speedup} for an individual generated program is calculated by averaging the speedups obtained across all the test cases.

\smallskip
The total amount of time required for conducting all the experiments --- including execution traces collection, model pre-training, fine-tuning and evaluation --- is approximately 3 weeks. The traces collection and evaluation procedures were conducted on a linux server equipped with 40 Intel(R) Xeon(R) 2.30GHz CPUs and 78Gb of RAM, while the models training and inference have been performed over a CentOS HPC cluster equipped with 32 Intel(R) Xeon(R) Gold 6140M CPUs and two Nvidia A100 and A30 GPUs.

\section{Results}\label{sec:results}
\begin{table*}[h!]
    \centering
    \scriptsize
    \caption{\emph{Correctness}, \emph{Speedup} and \emph{\%Opt} scores achieved by each baseline and proposed training strategy in performing code optimization of C++ programs. Scenarios that do not include pre-training have been reported in the bottom part of the table for a better understanding. For both the double-staged and single-staged strategies, the highest number in each column is \textbf{bolded} and the second-highest is \underline{underscored}.}
    \begin{tabular}{ll|ll|ccc}
    \toprule
    \multirow{2}*{\textbf{Model}} & \multirow{2}*{\textbf{\makecell{Execution Aspect}}} & \multicolumn{2}{c}{\textbf{Training Strategy}} & \multicolumn{3}{c}{\textbf{Evaluation Metrics}} \\ \cline{3-7}
     &  & \textbf{Pre-training} & \textbf{Fine-tuning} & \textbf{Correct} & \textbf{Speedup} & \textbf{\%Opt} \\
    \midrule
    $BL_{S_{12}}$ & - & - & Code optimization & \textbf{18.75\%} & \textbf{1.79} & \textbf{7.68\%} \\
    \midrule
    $LE_{S_{1}}$ & \multirow{2}{*}{\textit{Line Executions}} & Execution-aware & Code optimization & 12.36\% & 1.52 & 5.73\% \\
    $LE_{S_{2}}$ & & Execution-aware + MLM & Code optimization & 11.97\% & 1.54 & 5.6\% \\
    \midrule
    $LC_{S_{1}}$ & \multirow{2}{*}{\textit{Line Coverage}} & Execution-aware & Code optimization & 14.84\% & 1.7 & 6.9\% \\
    $LC_{S_{2}}$ & & Execution-aware + MLM & Code optimization & 14.45\% & \underline{1.76} & \underline{7.55\%} \\
    \midrule
    $BC_{S_{1}}$ & \multirow{2}{*}{\textit{Branch Coverage}} & Execution-aware & Code optimization & 13.93\% & 1.67 & 7.03\% \\
    $BC_{S_{2}}$ &                         & Execution-aware + MLM & Code optimization & 13.15\% & 1.55 & 5.73\% \\
    \midrule
    $VS_{S_{1}}$ & \multirow{2}{*}{\textit{Variable States}} & Execution-aware & Code optimization & \underline{15.49\%} & 1.69 & 7.29\% \\
    $VS_{S_{2}}$ & & Execution-aware + MLM & Code optimization &  14.97\% & 1.65 & 6.64\% \\
    \midrule
    \midrule
    $BL_{S_{3}}$ & - & - & Code optimization & \underline{12.06\%} & \underline{2.09} & \underline{9.22\%} \\
    \midrule
    $LE_{S_{3}}$ & \textit{Line Executions} & - & Execution-aware code optimization & \textbf{12.71\%} & \textbf{2.11} & \textbf{9.35\%} \\
    \midrule
    $LC_{S_{3}}$ & \textit{Line Coverage} & - & Execution-aware code optimization & 9.97\% & 1.67 & 5.84\% \\
    \midrule
    $BC_{S_{3}}$ & \textit{Branch Coverage} & - & Execution-aware code optimization & 8.65\% & 1.64 & 5.76\% \\
    \midrule
    $VS_{S_{3}}$ & \textit{Variable States} & - & Execution-aware code optimization & 11.55\% & 1.96 & 8.35\% \\
    \bottomrule
    \end{tabular}
    
    \label{tab:res_s1_s2_s3}
\end{table*}

Our evaluation follows a common methodology across all the research questions. For each execution-aware and baseline model, we first generate the optimized code and then compute the evaluation metrics defined in Section \ref{subsec:metrics}.
Table \ref{tab:res_s1_s2_s3} presents the results for the twelve execution-aware models and the two baselines. For the \emph{Speedup} metric, we report the mean \emph{Speedup} across all the programs of the test set.
The first column of the table specifies the execution-aware or baseline model being evaluated. For instance, $LE_{S_2}$ denotes the execution-aware model trained with line executions information ($LE$), using the strategy $S_2$.
$BL_{S_{12}}$ indicates the baseline model for strategies $S_1$ and $S_2$, while $BL_{S_3}$ denotes the baseline model for $S_3$.

Additionally, we assess the statistical significance of the differences between the speedup provided by the execution-aware model and baseline model using Wilcoxon's signed-rank test \cite{Wilcoxon1945}, and report two measures of effect size: Vargha-Delaney \vda \cite{Vargha2000} and the matched pairs rank biserial correlation \emph{r} \cite{Kerby2014}.
In this context, the \vda indicates the proportions of pairs where the \emph{speedup} is higher than the baseline. An \emph{\vda $>$ 0.5} indicates that a particular execution-aware model generates more efficient source code than the corresponding baseline. 
The matched pairs rank biserial correlation \emph{r} represents the difference between the proportion of favorable and unfavorable evidence. With regard to our work, favorable evidence is represented by a higher speedup for the execution-aware strategy, meaning that an \textit{r $>$ 0} is denoting that the execution-aware model performs better than the baseline.
Results from the comparison employing the Wilcoxon's signed-rank test, along with effect sizes, are reported in Table \ref{tab:wilc}. In the following, we discuss the achieved results by research questions.

\begin{table}[t]
    \scriptsize
    \center
    \captionsetup{justification=centering}
    \caption{Comparison of speedups achieved by execution-aware models vs. baselines, evaluated using Wilcoxon test, Vargha-Delaney \vda, and rank biserial correlation (\emph{r}).}
    \begin{tabular}{l|rrr}
        \toprule
        \textbf{Exec-aware \emph{vs.} Baseline} & \emph{p-}value & \textbf{\vda} & \emph{r} \\
        \midrule
        $LE_{S_{1}}$ \emph{vs.} $BL_{S_{12}}$ &              $<$0.05 &    0.446 (N) &     -0.643 \\
        $LE_{S_{2}}$ \emph{vs.} $BL_{S12}$ &              $<$0.05 & 0.467 (N) &     -0.399 \\
        $LE_{S_{3}}$ \emph{vs.} $BL_{S_3}$ &             0.064 &    - &     - \\ \midrule
        $LC_{S_{1}}$ \emph{vs.} $BL_{S_{12}}$ &              $<$0.05 & 0.451 (N) &     -0.633 \\
         $LC_{S_{2}}$ \emph{vs.} $BL_{S_{12}}$ &              $<$0.05 & 0.455 (N) &     -0.515 \\
         $LC_{S_{3}}$ \emph{vs.} $BL_{S_3}$ &              $<$0.05 & 0.471 (N) &     -0.416 \\ \midrule
        $BC_{S_{1}}$ \emph{vs.} $BL_{S_{12}}$ &              $<$0.05 & 0.452 (N) &     -0.536 \\
        $BC_{S_{2}}$ \emph{vs.} $BL_{S_{12}}$ &              $<$0.05 & 0.447 (N) &      -0.61 \\
         $BC_{S_{3}}$ \emph{vs.} $BL_{S_3}$ &              $<$0.05 & 0.47 (N) &     -0.503 \\ \midrule
        $VS_{S_{1}}$ \emph{vs.} $BL_{S_{12}}$ &             0.069 & - &      - \\
         $VS_{S_{2}}$ \emph{vs.} $BL_{S_{12}}$ &             0.075 & - &      - \\
         $VS_{S_{3}}$ \emph{vs.} $BL_{S_3}$ &              $<$0.05 & 0.487 (N) &     -0.449 \\
        \bottomrule
        \end{tabular}

\label{tab:wilc}
\end{table}

\subsection{RQ$_1$: How does learning line executions impact the effectiveness of language models for code optimization?}

By observing Table~\ref{tab:res_s1_s2_s3}, we find that both $LE_{S_1}$ and $LE_{S_2}$ models perform generally worse than the $BL_{S_{12}}$ baseline across all the evaluation metrics. 
Concerning the correctness of the generated code, we observe lower percentages compared to the baseline, with 12.36\% correct programs for  $LE_{S_1}$ and 11,97\% for $LE_{S_2}$, versus a correctness of 18.75\% for $BL_{S_{12}}$.
We also observe that execution-aware pre-training works slightly better for correctness when combined with masked language modeling (\ie $LE_{S_2}$). 
We find that the speedup achieved by the code generated by execution-aware models (1.52x and 1.54x for $LE_{S_1}$ and $LE_{S_2}$, respectively) is not as good as the one reached by the baseline (1,79x). The \emph{\%Opt} values confirm the same trend observed for the other two metrics, with lower percentages of optimized programs for the execution-aware models.

Contrariwise, we find that execution-aware fine-tuning ($LE_{S_3}$) works slightly better than its baseline $BL_{S_3}$ across all the evaluation metrics.
Particularly, $LE_{S_3}$ reports a higher number of correct programs (12.71\% vs. 12.06\%), a highest speedup (2.11x vs. 2.09x), and a higher percentage of optimized programs (9.35\% vs. 9.22). Since these improvements appear marginal, we further investigated the statistical significance of difference between the speedups provided by  $LE_{S_3}$ versus those of $BL_{S_3}$.
Table \ref{tab:wilc} shows the results of the Wilcoxon test, which suggest no statistical difference for the speedup, \ie  $p>0.05$.

\begin{tcolorbox}[size=title]
	\textbf{Answer to RQ$_1$:} 
Teaching line execution behavior to the language model does not enhance its code optimization capabilities.
Execution-aware fine-tuning shows a slight improvement, but without statistical significance.
\end{tcolorbox}

\subsection{RQ$_2$: How does learning line coverage impact the effectiveness of language models for code optimization?}

Results for line coverage reveal a trend similar to that observed in RQ1, with baselines outperforming all three execution-aware models.
When comparing $LC_{S_1}$ and $LC_{S_2}$, we observe an increase in speedup and \emph{\%Opt} values when the MLM is combined with execution-aware pre-training (1.76x vs. 1.7x for speedup, and 7.55x vs. 6.9x for \emph{\%Opt}), despite a slight reduction in correctness (14.45\% vs. 14.84\%).

By observing Table \ref{tab:res_s1_s2_s3}, we notice that, compared to other execution-aware models based on the strategies $S_1$ and $S_2$, line coverage demonstrates the highest effectiveness in terms of speedup and the percentage of optimized programs. 
This underlines that the effectiveness of execution-aware training strategies are closely related to the specific execution aspect used to feed the model.
For the execution-aware fine-tuning strategy $LC_{S_3}$, we observe a similar trend, showing lower effectiveness compared to the baseline $BL_{S_3}$, \eg speedup of 1.67x vs. 2.11x.

Wilcoxon's test indicates statistically significant differences for all comparisons between execution-aware models and baselines (p$<$0.05), with a negligible disadvantage for all execution-aware models in terms of effect size (\vda$<$0.5).

\begin{tcolorbox}[size=title]
	\textbf{Answer to RQ$_2$:} 
Learning line coverage does not improve the model effectiveness in optimizing code. 
Execution-aware models perform worse than baseline models across all metrics.
\end{tcolorbox}

\subsection{RQ$_3$: How does learning branch coverage impact the effectiveness of language models for code optimization?}

As shown in Table \ref{tab:res_s1_s2_s3}, execution-aware models consistently underperform compared to their respective baselines across all evaluation metrics. Specifically, we observe reduced correctness, with values of 13.93\%, 13.15\%, and 8.65\% for $BC_{S_1}$, $BC_{S_2}$, and $BC_{S_3}$, respectively. Regarding optimization metrics, execution-aware models also lag behind the baselines, achieving speed-ups of up to 1.67x (compared to 1.79x for $BL_{S_{12}}$ and 2.09x for $BL_{S_3}$) and \emph{\%Opt} values of 7.03\% (vs. 7.68\% for $BL_{S_{12}}$ and 9.22\% for $BL_{S_3}$).

When focusing exclusively on execution-aware fine-tuning, $BC_{S_3}$ exhibits worse effectiveness across all evaluation metrics compared to the baseline $BL_{S_3}$.

\begin{tcolorbox}[size=title]
	\textbf{Answer to RQ$_3$:}
The integration of branch coverage information into language models reduces both correctness and optimization metrics across the execution-aware training strategies.
\end{tcolorbox}

\subsection{RQ$_4$: How does learning variable states impact the effectiveness of language models for code optimization?}

The results for variable states in Table~\ref{tab:res_s1_s2_s3} indicate a degradation in all evaluation metrics when employing any of the considered execution-aware strategies. Specifically, correctness decreases on average from 18.75\% for the baseline $BL_{S_{12}}$ to 15.49\% for the execution-aware model $VS_{S_1}$ and further to 14.97\% for the $VS_{S_2}$ model. Regarding execution-aware fine-tuning, we observe a similar decline in correctness, from 12.06\% ($BL_{S_{12}}$) to 11.55\% ($VS_{S_3}$). Speedup and \emph{\%Opt} also follow a similar downward trend compared to the baselines.

\begin{tcolorbox}[size=title]
	\textbf{Answer to RQ$_4$:}
Learning variable states in language models does not enhance the effectiveness of code optimization. On the contrary, we observe reduced correctness and fewer optimizations in all execution-aware models.
\end{tcolorbox}

\section{Discussion}
\label{sec:discussion}

We conduct a complementary analysis to better understand the results of our study and to identify potential insights for future research on execution-aware code optimization.

\subsection{Correctness analysis}

\begin{table}[t]
    \scriptsize
    \center
    \captionsetup{justification=centering}
    \caption{
    Percentages of programs that successfully compile, execute, and produce correct outputs.}
    \begin{tabular}{l|ccc}
        \toprule
               \textbf{Model} &       \textbf{Compiled} &     \textbf{Executed} & \textbf{Correct} \\
        \midrule
        $BL_{S_{12}}$ &  78.65\% &   76.95\% &  \textbf{18.75\%} \\
        \midrule
        $LE_{S_{1}}$ &  81.25\% &   78.65\% &  12.36\% \\
        $LE_{S_{2}}$&   80.47\% &   77.21\% &  11.97\% \\
        \midrule
        $LC_{S_{1}}$ & 85.68\% &   82.94\% &  14.84\% \\
        $LC_{S_{2}}$ & 74.48\% &   72.53\% &  14.45\% \\
        \midrule
        $BC_{S_{1}}$ & 80.08\% &   77.08\% &  13.93\% \\
        $BC_{S_{2}}$ & \textbf{89.19\%} &   \textbf{86.72\%} &  13.15\% \\
        \midrule
        $VS_{S_{1}}$ &  85.03\% &   82.29\% &  15.49\% \\
        $VS_{S_{2}}$ & 85.68\% &   80.99\% &  14.97\% \\
        \midrule\midrule
        
        $BL_{S_3}$ &    \textbf{89.83\%} &   \textbf{88.65\%} &  12.06\% \\
        \midrule
        $LE_{S_{3}}$ &   86.81\% &   84.65\% &  \textbf{12.71\%} \\ \midrule
        $LC_{S_{3}}$ &   86.13\% &   84.18\% &   9.97\% \\ \midrule
        $BC_{S_{3}}$ &   85.10\% &   84.38\% &   8.65\% \\ \midrule
        $VS_{S_{3}}$ &   86.73\% &   84.77\% &  11.55\% \\
        \bottomrule
        \end{tabular}
    \vspace{10pt}
    
    \label{tab:corr}
\end{table}

Our evaluation reveals the limited capability of execution-aware models to produce correct code, \ie code that successfully passes all test cases. To better understand this limitation, we delve deeper into the reasons behind the lack of correctness. In particular, we analyze the proportions of generated programs that: (i) exhibit well-formed syntax (\ie, the code successfully compiles); (ii) do not result in runtime errors (\ie, the code successfully executes); and (iii) produce the expected output (\ie, the code passes all test cases)

Table~\ref{tab:corr} presents the percentages of programs that successfully compile (\emph{Compiled}), execute without errors (\emph{Executed}), and produce the expected output (\emph{Correct}).
We observe that execution-aware models exhibit comparable, and sometimes superior, effectiveness in generating programs that successfully compile and execute. Specifically, when considering execution-aware pre-training strategies ($S_1$ and $S_2$), these models outperform the baseline $BL_{S_{12}}$ in generating programs that compile and execute successfully.
For example, $LE_{S_1}$, $LC_{S_1}$, $BC_{S_2}$, and $VS_{S_2}$ generate programs that successfully compile in 81.25\%, 85.68\%, 89.19\%, and 85.68\% of cases, respectively, compared to 78.65\% for $BL_{S_{12}}$.
Similarly, $LE_{S_1}$, $LC_{S_1}$, $BC_{S_2}$, and $VS_{S_1}$ generate programs that successfully execute in 78.65\%, 82.94\%, 86.72\%, and 82.29\% of cases, respectively, compared to 76.95\% for $BL_{S_{12}}$.
However, when considering the correctness of the generated programs, we observe a notable drop in effectiveness for execution-aware models. Models using pre-training strategies $S_1$ and $S_2$ achieve a correctness of up to 15.49\%, whereas the baseline $BL_{S_{12}}$ achieves a correctness of 18.75\%. These results suggest that while execution-aware models generally perform well in generating well-formed code that successfully executes, they may be limited in their semantic understanding of the intended functionality (\ie correctness). This finding is somewhat counterintuitive, as execution-aware training strategies are typically designed to enhance language models' semantic understanding of code \cite{Ni2024,Ding2024SemCoder,Ding2024Traced}.
Nevertheless, in light of these findings, we encourage future research on code optimization to focus on improving the correctness of programs generated by execution-aware language models. This could involve exploring additional aspects of code execution or developing alternative, more effective training strategies.

\subsection{Speedup analysis}

\begin{table}[t]
    \scriptsize
    \captionsetup{justification=centering}
    \caption{
Speedup statistics considering only (i) correct programs and (ii) optimized programs.}
    \center\begin{tabular}{l|cc|cc}
        \toprule
        \multirow{3}*{\textbf{Model}} & \multicolumn{4}{c}{\textbf{Speedup}} \\
         & \multicolumn{2}{c|}{\textbf{Correct Programs}} & \multicolumn{2}{c}{\textbf{Optimized Programs}} \\
         &    \textit{\makecell{Instances}} &  \textit{Mean} &  \textit{\makecell{Instances}} &  \textit{Mean} \\
        \midrule
        $BL_{S_{12}}$ &                 144 &  5.2x &    141 & 5.29x \\
        \midrule
        $LE_{S_{1}}$ &  95 &  5.2x &     91 & 5.39x  \\
        $LC_{S_{1}}$ &           114 & 5.71x &     60 & \textbf{9.96x} \\
        $BC_{S_{1}}$ &     107 & 5.83x &      63 & 9.21x \\
        $VS_{S_{1}}$ &     119 & 5.44x &    111 & 5.76x  \\
        \midrule
        $LE_{S_{2}}$  &      92 & 5.49x  &     90 & 5.58x  \\
        $LC_{S_{2}}$  &     111 & \textbf{6.22x} &  66 & 9.79x  \\
        $BC_{S_{2}}$  &     101 & 5.21x  &     55 & 8.72x \\
        $VS_{S_{2}}$  &     115 & 5.37x  &    109 & 5.61x  \\
        \midrule\midrule
        $BL_{S_{3}}$  &      51 & \textbf{10.12x}  &     50 &  10.3x  \\ 
        \midrule
        $LE_{S_{3}}$ &      53 &  9.76x &     47 & 10.87x \\
        $LC_{S_{3}}$  &      41 &  7.76x  &     26 & \textbf{11.67x} \\
        $BC_{S_{3}}$  &      36 &   8.4x  &     26 & 11.24x \\
        $VS_{S_{3}}$  &      47 &  9.29x  &     40 & 10.74x  \\
        \bottomrule
        \end{tabular}
\label{tab:speedup}
\end{table}

We assess the average speedup across the entire set of generated code, accounting for both incorrect and slower programs. 
In line with prior research~\cite{shypula2024}, we assign a \emph{speedup = 1} in such cases.
To better understand the speedup provided by execution-aware models, here  we focus on analyzing only the correct programs generated by the models, as well as the programs that achieve an non-negligible optimization of execution time (greater than 1\%).
Table~\ref{tab:speedup} presents the mean speedup achieved by each model, calculated exclusively for programs that produce the expected outputs (\emph{Correct Programs}) and for those that optimize the input program (\emph{Optimized Programs}). Additionally, we report the number of program instances considered in each case. Execution-aware models deliver comparable or even superior speedup compared to baseline models when considering only correct programs. This trend is more evident in models utilizing execution-aware pre-training (\ie $S_1$ and $S_2$): $BC_{S_1}$ and $LC_{S_2}$ achieve mean speedups of 5.83x and 6.22x, respectively, whereas the baseline $BC_{S_{12}}$ achieves a speedup of 5.2x.
Furthermore, when focusing on optimized programs, we observe that execution-aware models achieve significantly higher speedups compared to baseline models. For instance, $LC_{S_1}$ achieves a speedup of 9.96x, while the baseline $BL_{S_{12}}$ 5.29x. Similarly, $LC_{S_3}$ achieves a speedup of 11.67x, compared to 10.3x for the baseline $BL_{S_3}$. These findings indicate that execution-aware models have the potential to deliver significant performance improvements when they successfully generate correct or optimized programs. 

\subsection{Threats to Validity}

\subsubsection{Construct validity}
Considering different execution aspects and quantization criteria could produce results that differ from those presented in this study.
Nevertheless, we focused on execution aspects that have been successfully employed in prior work \cite{Ding2024Traced}.
Using alternative training strategies to learn code execution information or larger language models may yield different outcomes.
Still, we evaluated a total of twelve execution-aware models by employing three distinct training strategies, and four execution aspects.
Additionally, we employed CodeT5+, a language model widely adopted in the software engineering literature \cite{Ahmed24,Luo24,Yu24,MastropaoloEASE24}.

\subsubsection{Internal validity}
The choice of the language model hyperparameters plays a key role in determining the model effectiveness. We try to mitigate possible biases by adopting the default parameters of the CodeT5+ release package. However, choosing different hyperparameters may lead to different results. Measuring program execution time typically involves a certain degree of variability \cite{Maricq2018}. To mitigate this risk and improve the scalability of our experiments, similar to previous work \cite{shypula2024}, we rely on a deterministic simulator, namely \emph{gem5}, to measure program execution time. However, executing programs on concrete hardware enviroments may yield different results.
In addition, the reliability of our performance benchmarking is significantly affected by the choice of the test cases, which we directly inherit from CodeNet.

\subsubsection{External validity}
Our evaluation is based on C++ programs from  CodeNet, which primarily consist of self-contained programs to solve specific coding problems. In addition, a random sampling was considered during dataset construction. The findings this study, they may not generalize to programs written in other programming languages or to more complex software systems involving multiple modules, third-party libraries, or intricate interdependencies.

\section{Conclusion}
\label{sec:conclusion}

We investigate the effectiveness of execution-aware language models in code optimization. We evaluate four execution aspects and three training strategies to teach a CodeT5+ how code executes at runtime. Our findings reveal that learning code execution behavior does not enhance the model's ability to optimize code.
A significant weakness of execution-aware models is their limited capacity to generate semantically correct code, even though they often produce syntactically correct programs. We encourage future research to explore additional code execution aspects, leverage other (larger) language models, and experiment with alternative training strategies.
This research objective informs and drives our future agenda.

\section*{Acknowledgements}

\newcommand{\rechargeAck}{Italian Government (Ministero dell'Università e della Ricerca, PRIN 2022 PNRR): ''RECHARGE: monitoRing, tEsting, and CHaracterization of performAnce Regressions`` (cod.P2022SELA7)\xspace}

\sloppy
The authors from University of L'Aquila are partially supported by \rechargeAck, and by ``ICSC – Centro Nazionale di Ricerca in High Performance Computing, Big Data and Quantum Computing'', funded by European Union – NextGenerationEU.
Bavota acknowledges the support from the Swiss National Science Foundation (SNSF), project “PARSED” (grant agreement No. 219294).


\balance

\bibliographystyle{IEEEtran}
\bibliography{references}

\end{document}